\documentclass[traditabstract]{aa} 
\usepackage{graphicx}
\usepackage{natbib}
\usepackage{mathabx}
\usepackage{deluxetable}
\usepackage[pdftex,                
bookmarks=true,                    
bookmarksnumbered=true,            
colorlinks=true,            
citecolor=blue,         
linkcolor=blue,     
menucolor=blue,         
urlcolor=cyan,       
linkbordercolor={0 0 1},           
pdfborder= {0 0 1},
frenchlinks=True]{hyperref}
\usepackage{txfonts}
\providecommand{\eprint}[1]{\href{http://arxiv.org/abs/#1}{#1}}
\providecommand{\adsurl}[1]{\href{#1}{ADS}}

\def\micron{\ensuremath{\mu m}}

\def\vsini{\ensuremath{v \sin i}}
\def\kms{\ensuremath{\textrm{km~s}^{-1}}}

\newcommand{\reff}{}

\begin{document}

\title{Doppler Imaging of Exoplanets and Brown Dwarfs}

\authorrunning{Crossfield}
\titlerunning{Doppler Imaging of Exoplanets and Brown Dwarfs}

   \author{Ian J. M. Crossfield
             }

          \institute{
             Max-Planck Institut f\"ur Astronomie, K\"onigstuhl 17, 69117, Heidelberg, Germany \\
            \email{\href{mailto:ianc@mpia.de}{ianc@mpia.de}}
            }

   \date{Submitted to A\&A: 2014 March 3. Resubmitted: 2014 April 28.}

 
  \abstract
  {

    Doppler Imaging produces 2D global maps of rotating objects using
    high-dispersion spectroscopy. When applied to brown dwarfs and
    extrasolar planets, this technique {\reff can} constrain global
    atmospheric {\reff dynamics and/or magnetic effects on} these
    objects in unprecedented detail. I present the first quantitative
    assessment of the prospects for Doppler Imaging of substellar
    objects with current facilities and with future giant ground-based
    telescopes. Observations will have the greatest sensitivity in K
    band, but the H and L bands will also be useful for these
    purposes.  To assess the number and availability of targets, I
    also present a compilation of all measurements of photometric
    variability, rotation period ($P$), and projected rotational
    velocity (\vsini) for all known brown dwarfs. Several bright
    objects are already accessible to Doppler Imaging with currently
    available instruments. With the development of giant ground-based
    telescopes, Doppler Imaging will become feasible for many dozens
    of brown dwarfs and for the few brightest directly imaged
    extrasolar planets (such as $\beta$~Pic~b). The present set of
    measurements of $P$, \vsini, and variability are incomplete for
    many objects, and the sample is strongly biased toward early-type
    objects ($<L5$).  Thus, surveys to measure these quantities for
    later-type objects will be especially helpful in expanding the
    sample of candidates for global weather monitoring via Doppler
    Imaging.

}
   { }
   { }
   { }
   {}


   \maketitle
%

\section{Introduction}

The use of high-dispersion, near-infrared (NIR) spectrographs is
quickly advancing the study of brown dwarfs and extrasolar planets.
Work in recent years has revealed the atmospheric composition and
thermal structure of numerous short-period exoplanets
\citep[e.g.,][]{snellen:2010,crossfield:2011,brogi:2012,rodler:2012},
the projected rotational velocity of the young exoplanet $\beta$~Pic~b
\citep{snellen:2014}, and has produced the first 2D global map of
patchy clouds on a brown dwarf \citep{crossfield:2014a}. With a
rapidly growing number of instruments suited to this work (see
Table~\ref{tab:instruments}), the stage is set for exciting new
developments in the fields of planetary and brown dwarf science.

The Doppler Imaging map of the nearby brown dwarf Luhman~16B
demonstrates the first opportunity to study global atmospheric
dynamics and circulation and two-dimensional surface brightness
variations -- i.e., weather patterns -- on {\reff planets and brown
  dwarfs} beyond the Solar system. Photometric variability has been
observed on brown dwarfs for many years \citep{tinney:1999}, with
{\reff variability seen across all ages and spectral types}
\citep{joergens:2003,buenzli:2014}.  Disk-integrated
measurements of variability provide many significant insights into our
understanding of brown dwarf atmospheres {\reff both in young
  star-forming regions \citep{joergens:2003,cody:2010} and for older
  objects in the field}
\citep{artigau:2009,radigan:2012,buenzli:2012,apai:2013,heinze:2013}
and just as they do for hot Jupiters \citep[e.g.,][]{knutson:2012}.  But
degeneracies inevitably {\reff preclude any unique determination of}
the brightness and geographic distributions  of surface features
from disk-integrated {\reff photometry} \citep{cowan:2008,cowan:2013,apai:2013}.

\begin{figure*}[bt!]
\centering
\includegraphics[width=7.5in]{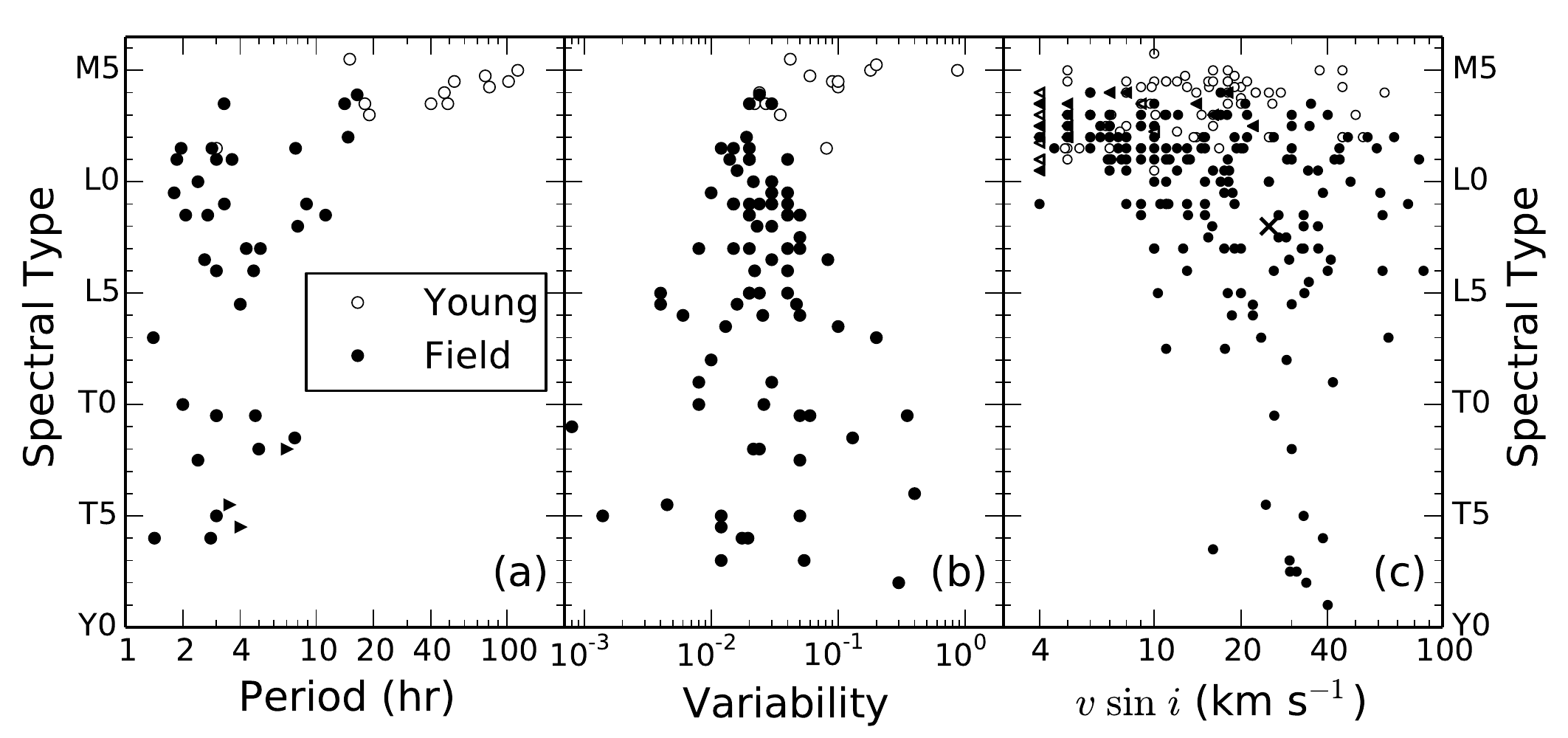}
\vspace{-0.3cm}
\caption{ \label{fig:prv} Potential targets for future Doppler
  Imaging: the full catalogue of measurements of substellar rotation
  and variability.  The three panels show the spectral types of
  objects with measured rotation periods (a), photometric variability
  semi-amplitude in any bandpass (b), and \vsini\ (c).  Upper limits
  to \vsini, and lower limits on periods, are indicated by
  triangles. {\reff Solid points are field objects, empty points are
    young objects. The '$\times$' symbol in panel (c) indicates the
    young, giant exoplanet $\beta$~Pic~b \citep{snellen:2014}.} See
  Table~\ref{tab:database} for the full compilation and
  references. Note that the sample is quite incomplete for objects
  later than $\sim$L5.}
\end{figure*}

Doppler Imaging can break these degeneracies.  With appropriate
regularization, the method produces a unique two-dimensional surface
map of the entire visible surface of a rotating body.  The technique
dates back at least to \cite{deutsch:1958} and has been recast and
reviewed multiple times in the succeeding years
\citep{vogt:1987,piskunov:1993,rice:2002}. In brief, the approach
takes high S/N rotationally-broadened line profiles and inverts
modulations of these lines (induced by heterogeneous regions of
differing abundances and/or surface brightness) into a global map
using matrix-based or iterative forward-modeling approaches.

Doppler Imaging of brown dwarfs is an exciting new development because
in many ways these objects provide a direct proxy for studying
planetary atmospheres.  Previous Doppler Imaging studies were limited
to cool stars whose variability was traced to persistent or
slowly-evolving magnetic starspots
\citep[e.g.,][]{cameron:1995,hatzes:1998,jardine:1999,strassmeier:2009},
{\reff and the high-amplitude variability seen in young brown dwarfs
  likely results from similar processes \citep{joergens:2003}.}
Cooler, {\reff field} brown dwarfs have essentially neutral
photospheres and so magnetic effects are {\reff likely} less
important.  Instead, variability in these objects is {\reff thought to
  be} dominated by effects such as variable cloud thickness
\citep{burgasser:2002,heinze:2013} and global circulation and
convection \citep{freytag:2011,showman:2013, zhang:2014} -- the same
processes that dominate the observable atmospheres of Jupiter and the
other Solar system gas giants.

In this paper I present the first quantitative analysis of the
prospects for Doppler Imaging of brown dwarfs and of directly imaged
giant extrasolar planets. These analyses indicate that while a few
brown dwarfs are amenable to mapping with existing facilities, Doppler
Imaging will be feasible for dozens or hundreds of objects -- both
planets and brown dwarfs -- with planned giant ground-based telescopes
such as GMT, TMT, or E-ELT {\reff
  \citep{johns:2008,nelson:2008,gilmozzi:2007}.}

I also present a compilation of the rotational periods, photometric
variability, and/or projected rotational velocities reported for all
{\reff brown dwarfs and exoplanets}.  All of these quantities must be
known before applying Doppler Imaging to a target, and so the included
catalog represents a first step toward a target list for future
Doppler Imaging efforts. In addition, such a catalog is useful for
studies of, e.g., angular momentum, cloud properties, and physical
sizes of substellar objects.

Sec.~\ref{sec:database} describes the new catalog of measurements of
variable and rotating substellar objects. Sec~\ref{sec:sens} presents
scaling relations for estimating the sensitivity of Doppler Imaging
efforts and estimates limiting magnitudes for various combinations of
spectral type and telescope aperture. Finally,
Sec.~\ref{sec:conclusion} concludes with a summary and final thoughts.

\section{A New Catalog of Variable and Rotating Substellar Objects}
\label{sec:database}

\subsection{Compiling the Catalog}
To begin evaluating targets for Doppler Imaging with the instruments
summarized in Table~\ref{tab:instruments}, I have compiled a new
database that contains all brown dwarfs and exoplanets for which
published measurements exist of photometric variability, rotation
periods ($P$), and projected rotational velocities (\vsini).
Fig.~\ref{fig:prv} plots measurements of $P$, \vsini, and photometric
variability against spectral type. The catalog data are
presented in Table~\ref{tab:database} and the full data set is
available online in machine-readable format.

In total the catalog contains {\reff 347} objects: {\reff 58} with
reported rotation periods, {\reff 109} exhibiting variability in
optical or infrared photometric bands, and {\reff 271} with measured
\vsini.  For {\reff 23} objects, all three quantities have been
measured. The electronic catalog includes spectral types (from
\url{http://www.dwarfarchives.org} and the literature) and $JHK_s$
magnitudes from 2MASS \citep{skrutskie:2006}.  The full set of
references for all entries are included in the catalog and cited in
the Appendix.

When an object's \vsini\ was measured by multiple groups, I report the
weighted mean value; upper limits are also included. For rotation
periods, I do not include the uncertainty on $P$ which is often
difficult to quantify and only infrequently reported.  Variability
measurements in the literature are more inhomogeneous in quality than
are either $P$ or \vsini. Despite some concern regarding variability
detections of questionable significance, recent observations
demonstrate that for some brown dwarfs the amplitude of variability
can change considerably from one epoch to the next
\citep[e.g.,][]{gillon:2013}. I therefore liberally include all
significant reports of previous variability, but I do not include null
detections.

Fig.~\ref{fig:prv} demonstrates that the catalog is dominated by
early-type objects.  Most measurements have been made for objects with
earlier spectral types ($\lesssim$L5) because these intrinsically
brighter objects are easier to detect and study at high signal to
noise (S/N). Aside from the recently-discovered and well-studied T0.5
Luhman~16B \citep{luhman:2013}, the coolest object with measured $P$,
\vsini, and photometric variability has spectral type L4. Much work
remains to be done, then, in completely characterizing the late-type
sample.

\subsection{Rotational velocities and periods}
\label{sec:rot}

\begin{figure}[bt!]
\centering
\includegraphics[width=3.5in]{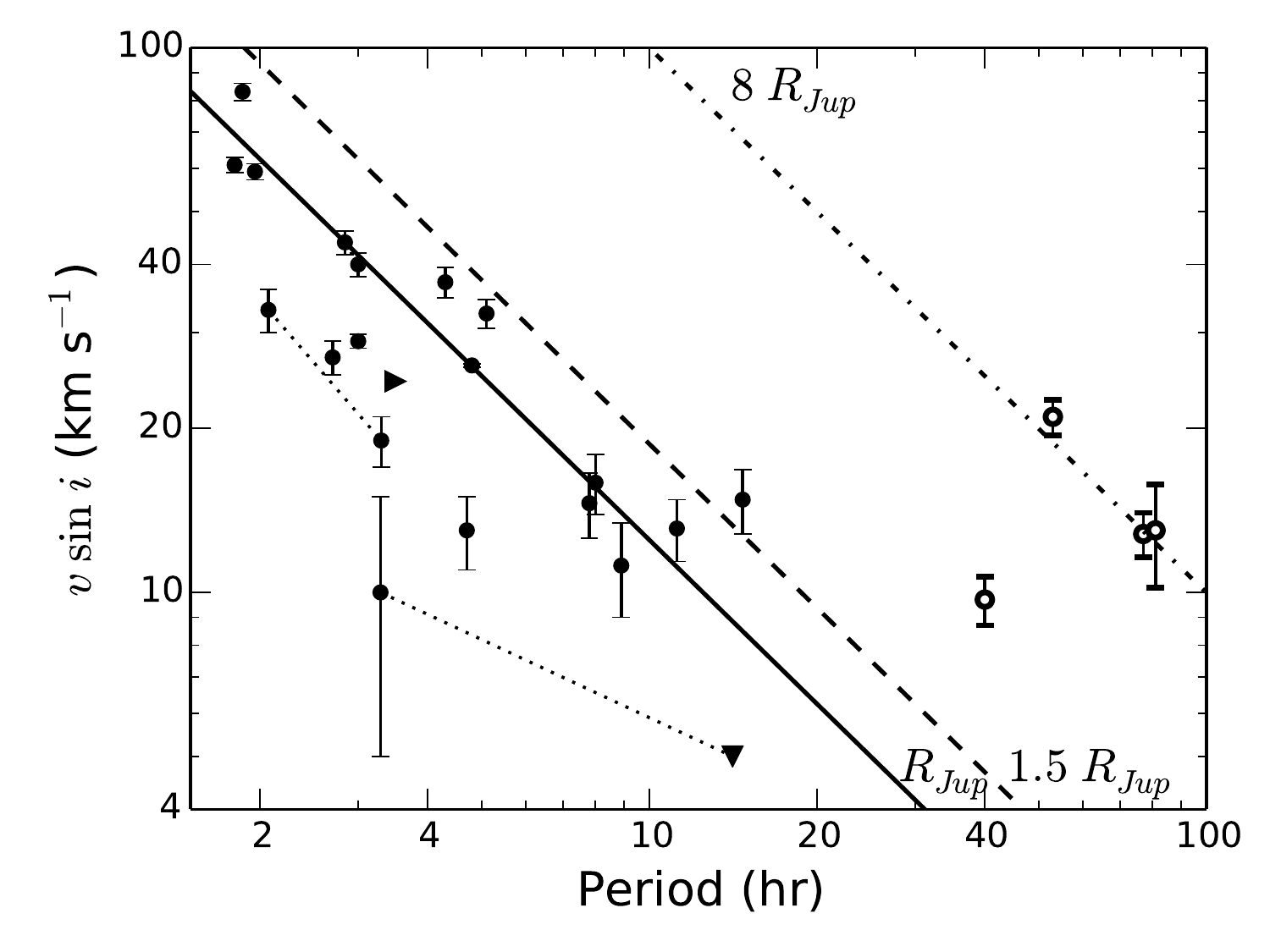}
\vspace{-0.3cm}
\caption{ \label{fig:pv} Brown dwarfs with measured rotation periods
  and \vsini. The curves show the maximum possible \vsini\ for the
  indicated radii. Open points are young substellar objects in the
  $\sigma$~Ori or Cha~I star forming regions. Dotted lines connect the
  components of binary systems. The triangles indicate upper or lower
  limits. }
\end{figure}

As one example of the utility of such a database, Fig.~\ref{fig:pv}
shows {\reff the 23} brown dwarfs with measured $P$ and \vsini. For
known $P$, the upper limit to \vsini\ is $2\pi R/P$.  For a radius of
$R_{Jup}$, this limit is indicated by the solid line in
Fig.~\ref{fig:pv}. {\reff Measurements indicate nine objects with
  radii $>R_{Jup}$: four objects in young star forming regions, and
  five field objects.} In order of increasing $P$, these objects are:
field objects LP~349-25B, DENIS~1058-1548, 2MASS 1146+2230AB, 2MASS
1146+2317, and LP~412-31; {\reff and young objects S~Ori~25, and
  Cha~H$\alpha$~3, 2, and~6.}

Previous analyses of these objects are consistent with the conclusion
that they must have radii $>R_{Jup}$.  \cite{harding:2013b} estimate
the radius of LP~349-25B to be 1.45~$R_{Jup}$, consistent with
Fig.~\ref{fig:pv}.  \cite{heinze:2013} discuss the tension between
various pieces of evidence in reconciling the rotational and physical
parameters of DENIS~1058-1548, but favor a radius
$>1.1R_{Jup}$. 2MASS~1146+2230AB is a close (0.3'') binary and it is
not clear which component the literature values for $P$ and \vsini\
refer to; nonetheless, the large $\vsini = 32.5$~\kms\ must be
dominated by rotational effects. \cite{bailerjones:2004} note that $P$
and \vsini\ of both 2MASS~1146+2230AB and 2MASS~1146+2317 can be
reconciled if $R\ge1.3 R_{Jup}$ and $\ge 0.95 R_{Jup}$, respectively,
consistent with Fig.~\ref{fig:pv}.  \cite{irwin:2011} note that
LP~412-31 must have $R > 1.1 R_{Jup}$, consistent with
Fig.~\ref{fig:pv}.  {\reff Measurements for the young substellar
  objects shown in Fig.~\ref{fig:pv} are also} self-consistent because
these objects are expected to be quite large ($>7 R_{Jup}$) on account
of their young age \citep{joergens:2003, caballero:2004}.

In addition to these systems, {\reff 35} brown dwarfs have  rotation
periods but no reported \vsini. High-resolution spectroscopy would
more than double the current ($P$, \vsini) sample, providing useful
insights into brown dwarf evolution and (via gravity-sensitive
features) help constrain absolute masses and radii of these objects.

\vspace{1in}
\section{Estimating Sensitivity for Doppler Imaging}
\label{sec:sens}
%
Advances in Doppler Imaging of substellar objects will rely primarily
on improved instruments and telescopes.  The first Doppler Image of a
brown dwarf was produced by observing an exceptional object -- a
2~pc-distant T0.5 dwarf with $K_s$=9.8~mag -- with VLT/CRIRES, which
offers high dispersion but a wavelength coverage of only 60~nm
\citep[][]{crossfield:2014a}. Aside from a small number of brown
dwarfs near the M/L transition, few other substellar objects are
likely to be discovered that are so bright or so nearby. Yet
Table~\ref{tab:instruments} demonstrates that many spectrographs
offering much broader wavelength coverage than CRIRES are currently
under development.

\subsection{A Rough Sensitivity Metric}
As already discussed, successful Doppler Imaging relies on precise
measurements of the changing shapes of rotationally broadened
absorption line profiles. One way to boost the precision is to
increase the number of lines -- and thus the total equivalent width
(EW) -- monitored in the analysis, either through techniques such as
least-squares deconvolution \citep{donati:1997} or through a full
spectral synthesis approach. Precision will also increase when one
obtains an intrinsically higher spectroscopic S/N -- whether by
observing a brighter target, using a larger telescope, or building a
more efficient spectrograph. Finally, when all else is equal
higher-fidelity maps can be made for objects exhibiting higher surface
brightness heterogeneities; these are the features that manifest
themselves as photometric variations with semi-amplitude $\Delta F$.

Combining these factors produces a metric that describes how suitable
a given target is for Doppler Imaging.  This {\reff ``mapping
  sensitivity''} metric is
\begin{equation}
\label{eq:metric}
M =  \Delta F  \times \left( S/N \right) \times  \sum EW  / 2.5\micron
\end{equation}
The denominator is chosen to yield $M\approx1.5$ for the initial
Doppler Imaging analysis of Luhman 16B, which provides a map that is
useful but still rather noisy compared to analyses of brighter,
stellar targets.  I therefore suggest a cutoff of $M>1$ as a rough
rule of thumb. Objects falling below this threshold will yield only
low-quality maps owing to being too faint, having too few strong
absorption features, and/or having too nearly featureless a surface.
{\reff The spectra of very late-type objects exhibit a nearly
  uninterrupted forest of often-weak absorption lines; a sufficiently
  high \vsini\ may blur these lines to such an extent that Doppler
  Imaging is no longer feasible. The impact of this effect should be
  studied in the future, but the results below demonstrate that in the
  near term Doppler Imaging will likely only be applied to objects
  with spectral types earlier than Luhman~16B.}

Naturally, other factors will also influence the fidelity of a Doppler
Imaging analysis.  For example, spectral resolution ($R$) limits the
number of longitudinal (spatial) resolution elements around the
resulting map to be roughly $n \le R \vsini / c$ (N.~Piskunov, private
communication). When observing over a single rotation period, the
period $P$ sets a similar limit of $n \le P/t_{int}$, where $t_{int}$
is the effective integration time.  Furthermore, the evolution of
photometric variability varies considerably for different objects,
ranging from short-term variations with little periodicity to strongly
periodic signatures that repeat for many rotations: the latter objects
in particular could be observed over many rotations to increase the
quality of the desired Doppler Image.  In practice, a full simulation
should be run for any target on the basis all known parameters:
spectral morphology, brightness, rotation period, \vsini, and
amplitude and type of photometric variability. Nonetheless
Eq.~\ref{eq:metric} is useful because it allows quick and simple
evaluation of large numbers of targets.

\subsection{Simulated Observations}

\begin{figure*}[bt!]
\centering
\includegraphics[width=7.5in]{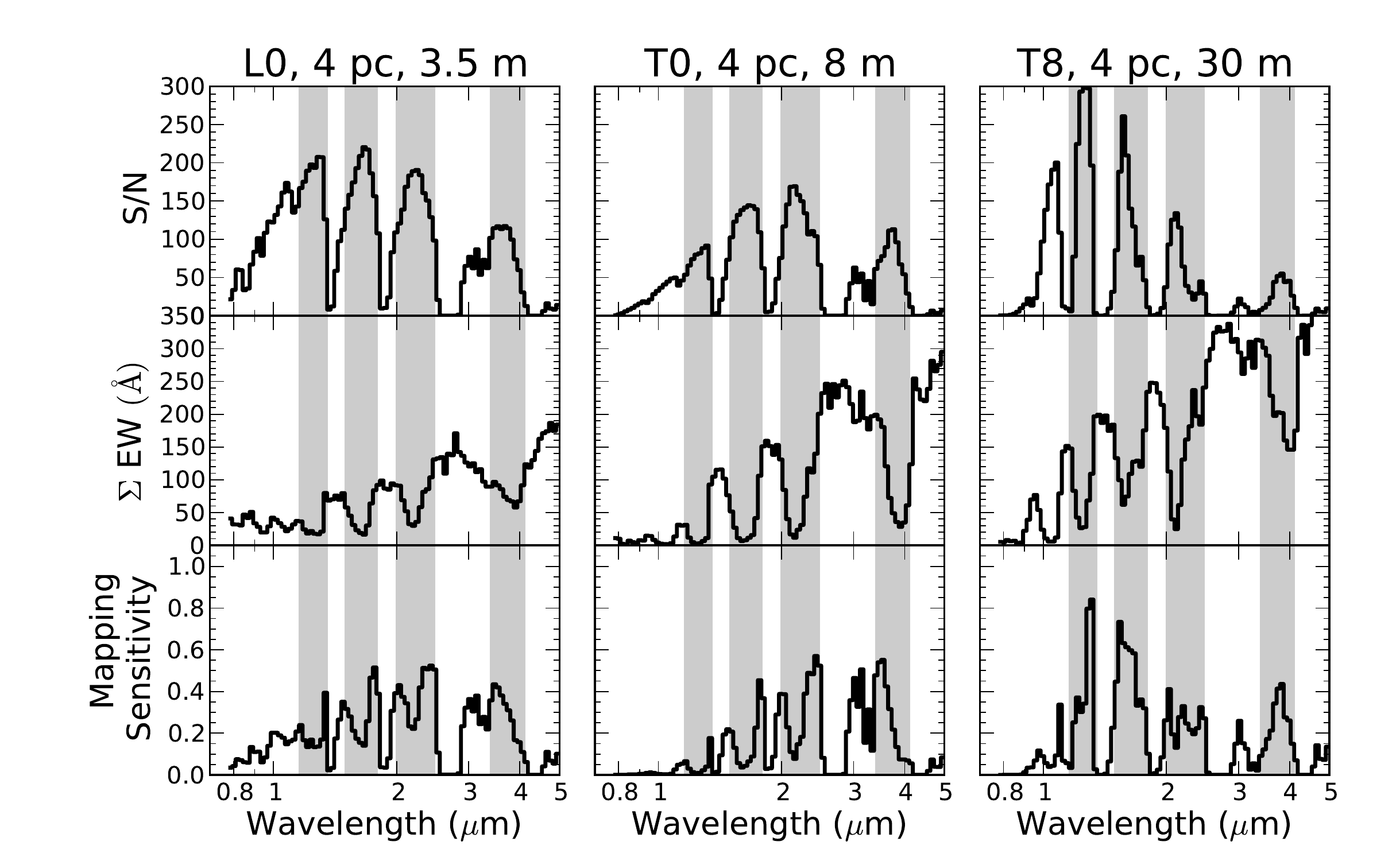}
\vspace{-0.3cm}
\caption{ \label{fig:metrics} Doppler Imaging sensitivity analyses
  for three representative cases: an L0 dwarf observed with a 3.5~m
  telescope (left), a T0 with an 8~m telescope (center), and a T8 with
  a 30~m telescope (right), all located 4~pc away. From bottom to top,
  the plots show (in $0.02\lambda$ bins) the median S/N, total
  equivalent width of absorption lines, and the normalized product of these
  quantities . The root sum square of this last quantity over
  observable wavelengths determines the final Doppler Imaging
  sensitivity of a given instrument. The gray regions indicate the
  wavelength regions used to calculate the magnitude limits shown in
  Fig.~\ref{fig:sens}. }
\end{figure*}

I next use the {\reff mapping sensitivity} metric to estimate the feasibility
of conducting Doppler Imaging for various types of brown dwarfs using
several representative observing facilities.  This effort requires two
components: a large set of brown dwarf spectra, and a simple model of
a high-dispersion spectrograph.

Spectra were taken from the BT-Settl library, which provides
high-resolution model surface fluxes for objects with effective
temperatures from 400~K to 2600~K \citep{allard:2014}. All
models had surface gravities of $10^5 \textrm{~cm~s}^{-2}$ and used the
abundance ratios of \cite{caffau:2011}.

For the instrument, I developed a simple radiometric model of a
high-dispersion NIR spectrograph. I use plausible values for the
necessary instrument parameters, as compared to the various
instruments listed in Table~\ref{tab:instruments}. The sky emission is
appropriate for Mauna Kea and comes from the Gemini Observatory
website\footnote{Currently at \url{
    http://www.gemini.edu/sciops/telescopes-and-sites/observing-condition-constraints/ir-background-spectra}
  ; data there are based on the ATRAN model of \cite{lord:1992}.}. I
assume a slit width of 0.2'', a pixel scale of 0.1''~pixel$^{-1}$ and
$10^{-5}\,\lambda\, \textrm{pixel}^{-1}$, {\reff a spectral resolution
  of $5\times10^4$,} and a total instrumental throughput of
10\%. Telluric transmission is taken from the high-resolution
measurements of \cite{hinkle:2003}. Each simulated spectrum is
interpolated onto 2048~pixels spanning a wavelength range of
0.02$\lambda$, which approximates one echelle order {\reff and gives
  $\sim$10 orders per telluric window.} Finally, each simulated
integration assumes 30~min of observing time: considering that brown
dwarfs have rotation periods as short as 2~hr (see Figs.~\ref{fig:prv}
and~\ref{fig:pv}), this integration time seems the longest likely to
provide useful longitudinal resolution in a Doppler Imaging analysis.

With this tool in hand, I compute the median S/N and the total EW of
all absorption lines in each echelle order. The continuum level for
computing EW values is determined by fitting a low-order polynomial to
the brown dwarf's spectrum and iteratively rejecting points
significantly below the fit. The S/N, total EW, and the {\reff mapping sensitivity}
(the product of S/N and EW) are plotted in Fig.~\ref{fig:metrics} for
three representative cases. To compute $M$ I assume a photometric
variability of $\Delta F=3\%$, which is the median of the values in
Table~\ref{tab:database} and Fig.~\ref{fig:prv}b.

To capture the benefits of the broad wavelength coverage offered by
modern spectrographs, I compute sensitivity limits assuming a Doppler
Imaging analysis that uses a single photometric band. To the extent
that many instruments listed in Table~\ref{tab:instruments} can
observe multiple bands simultaneously, this assumption underestimates
the true sensitivity for a given target. The effective, broadband
{\reff mapping sensitivity} ($M_{bb}$) is then the root-sum-square of
all values of $M$ computed as described above that lie within the
specified bandpass. The limits I use are indicated by the gray regions
in Fig.~\ref{fig:metrics}, and are: $1.16-1.35\,\micron$,
$1.5-1.79\,\micron$, $2.0-2.47\,\micron$, and $3.4-4.1\,\micron$. From
the definition of $M$ in Eq.~\ref{eq:metric}, a value of $M_{bb}\ge1$
for a given simulation indicates that Doppler Imaging is feasible for
the given combination of telescope, target effective temperature, and
distance.

\begin{figure*}[bt!]
\centering
\includegraphics[width=7.5in]{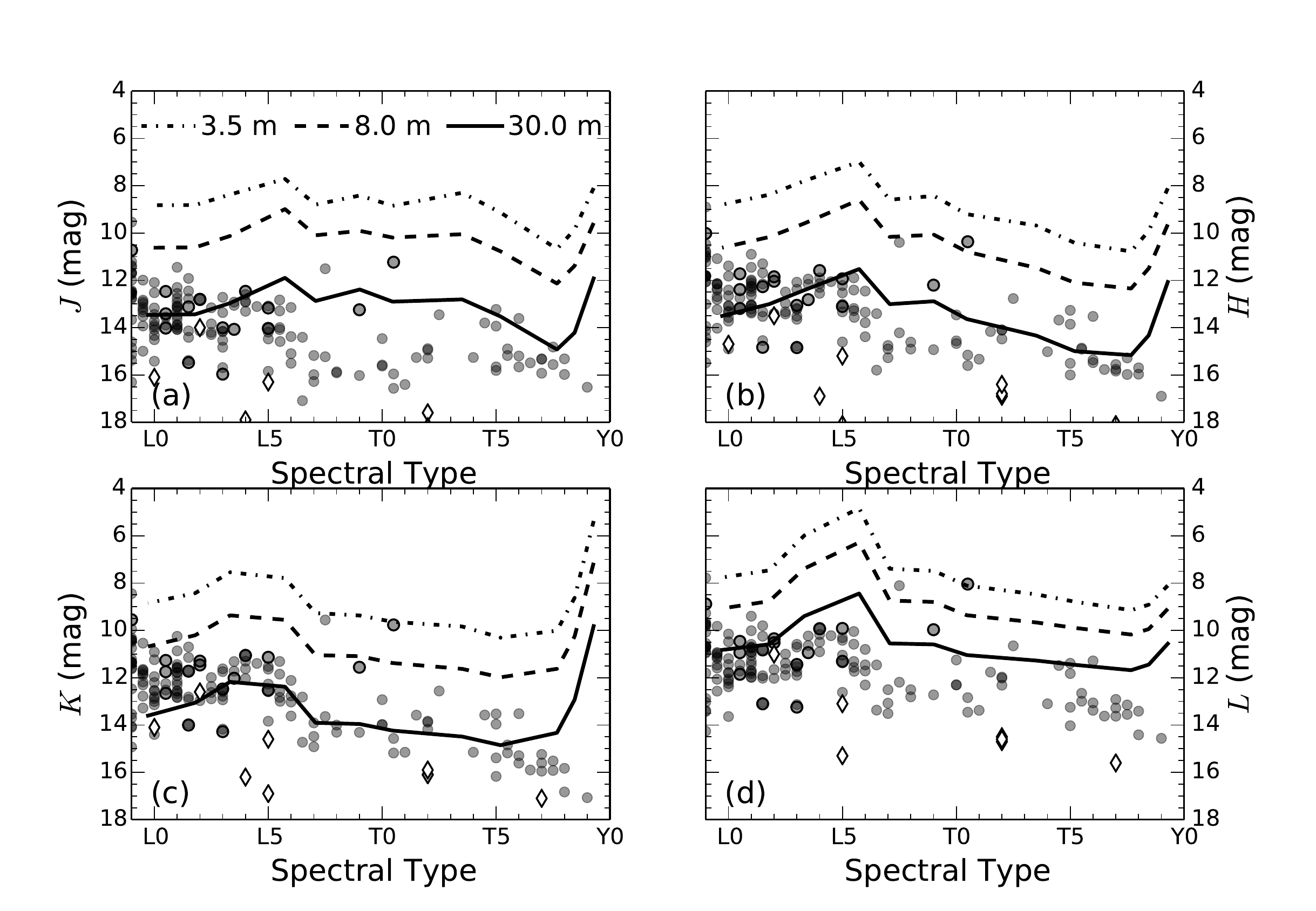}
\vspace{-0.3cm}
\caption{ \label{fig:sens} Limiting infrared magnitudes for NIR
  Doppler Imaging. Panels (a)--(d) show magnitude vs.\ spectral type
  for J, H, K, and L bands for a telescope either 3.5~m (dash-dotted),
  8~m (dashed), or 30~m (solid) in diameter. Points indicate objects
  with measured rotational periods, photometric variability, and/or
  \vsini. The best potential targets for Doppler Imaging are indicated
  with black outlines, which have $\vsini \ge15\ \kms$ and/or vary
  photometrically by $\ge 2\%$. As described in Sec.~\ref{sec:sens}
  these limits assume Doppler Imaging in only a single NIR band;
  observing slightly fainter objects should be feasible with broader
  wavelength coverage.  Rhombus symbols indicate known directly-imaged
  giant exoplanets: a small number of these may be amenable to Doppler
  Imaging with future large-aperture telescopes.}
\end{figure*}

All these calculations provide an estimate of the greatest distance --
or equivalently, the faintest apparent magnitude -- an object could
have and still be a useful target for Doppler imaging. These results
are shown in Fig.~\ref{fig:sens} for three different telescope
apertures (3.5~m, 8\,m, and 30\,m) for observations in each of the J,
H, K, and L bandpasses. To make this plot, I converted the effective
temperatures of the BT-Settl models to spectral type using the
relation of \cite{stephens:2009}.

In addition to the sensitivity limits, Fig.~\ref{fig:sens} shows the
apparent magnitudes and spectral types of all substellar objects
included in the catalog presented above. I approximate the L band
magnitudes using the relation $K-L=0.092 T - 0.16$ \citep[Figure 1
of][]{golimowski:2004}, where $T$ is the spectral type: 0 for M0, 10
for L0, etc. The scatter on this relationship is $\lesssim0.2$~mag, as
estimated from the fits of spectral type vs. absolute magnitude
relations for K band \citep{looper:2008} and L band
\citep{golimowski:2004}.

\subsection{Results: Feasibility of Substellar Doppler Imaging}
\label{sec:results}
Fig.~\ref{fig:sens} indicates that K band is the optimal wavelength
range for Doppler Imaging of brown dwarfs. A high-dispersion
spectrograph covering the K band could image an object like Luhman~16
-- 2~pc distant, at the L/T transition -- even on a $<4$\,m telescope
(see Fig.~\ref{fig:sens}c).  In contrast, the prospects at J band are
the poorest of the wavelength ranges considered: Fig.~\ref{fig:sens}a
shows that Doppler Imaging at these wavelengths will likely be
feasible only with future giant ground-based telescopes.  {\reff These
  results should not change so long as the amplitudes of brown dwarf
  variability does not hugely vary for these different bandpasses. For
  example}, Fig.~\ref{fig:sens} {\reff shows that for a given
  brightness, an L/T transition dwarf would have to vary $\sim4\times$
  more at H~band than at K~band to provide comparable Doppler Imaging
  sensitivity in both bandpasses. Multiwavelength observations
  demonstrate that the variability amplitudes of these objects are
  constant to within a factor of $\sim$2 in different, broad
  bandpasses} \citep{radigan:2012,buenzli:2012,biller:2013}.

In general, 3-4\,m telescopes will be limited to the brightest objects
at any given spectral type. Nonetheless Doppler Imaging of several new
brown dwarfs, especially the brightest targets near the M/L
transition, should be feasible with these instruments. One example is
the 4~pc-distant M9 dwarf 2MASS~1048147-395606, which has
$\vsini=18\pm2\,\kms$ but no detections of photometric variability or
rotation period.

Doppler Imaging of more targets will become feasible with new,
broad-wavelength-coverage spectrographs on 8\,m telescopes. A few more
later-type objects could be observed at K and/or L band. These include
the T1 dwarf 2MASS~J22431696-5932206 (for which 3\% variability has
been reported at I band, but $P$ and \vsini\ have not been measured)
and the rapidly-rotating L9 object DENIS~0255-4700 (which has $\Delta
F=3\%$ and $\vsini=41.7 \pm 1.5\,\kms$, but no measured period).

With $\sim$30\,m-class telescopes, Fig.~\ref{fig:sens} demonstrates
that Doppler Imaging will become feasible for many dozens of
substellar objects. Many of these will lie near the M/L transition,
but at spectral types later than L5 at least 10--15 plausible
candidates for these future studies already exist. As noted
previously, measurements of photometric variability and \vsini\ are
much less complete for later-type objects: thus many potential targets
probably remain to be found.

Perhaps even more exciting, with these large new telescopes Doppler
Imaging of a small number of extrasolar giant planets may be
possible. That this is plausible should not be too surprising, since HR~8799c has
already been detected at good S/N using medium-resolution K band
spectroscopy \citep{konopacky:2013}. The rhombus symbols in
Fig.~\ref{fig:sens} indicate the magnitudes and approximate spectral
types of several of the brightest known extrasolar planets
\citep{chauvin:2004,chauvin:2005,marois:2008,marois:2010,lafreniere:2008,barman:2011,carson:2013,bonnefoy:2013,esposito:2013}. Of
all these objects, the L2-type planet $\beta$~Pic~b seems the best
candidate for Doppler Imaging, with AB~Pic~b (L0) another possible
target. These estimates are only approximate, since the spectral
energy distributions of young, self-luminous planets differ from those
of the brown dwarf spectra used in Sec.~\ref{sec:sens}. Though
variability and rotation parameters have not yet been reported for
directly imaged planets, the analysis presented here suggests that
such measurements -- and eventually, Doppler Imaging -- might be
feasible for these low-mass objects.

These estimated limits must of course be modulated by the level of
photometric variability, which is not directly encapsulated by
Fig.~\ref{fig:sens}. As noted above, the sensitivity curves shown
correspond to a photometric semi-amplitude of 3\%, the median of the
sample shown in Fig.~\ref{fig:prv}b. Based on Eq.~\ref{eq:metric}, for
an object with variability $f$ times this level the sensitivity curves
in Fig.~\ref{fig:sens} should be shifted by $5\log_{10}
f$. Nonetheless, the trends described above will not change to the
extent that variability amplitude and frequency are independent of
spectral type \citep{buenzli:2014}.

\section{Discussion, Conclusions, and Future Work}
\label{sec:conclusion}
I have presented the first quantitative investigation of the
feasibility of Doppler Imaging of brown dwarfs and extrasolar
planets. The main results are summarized in Fig.~\ref{fig:sens}, in
terms of approximate limiting magnitudes for high-dispersion infrared
spectrographs on 3.5\,m, 8\,m, and 30\,m telescopes.  K band offers
the best prospects for Doppler Imaging of objects with spectral types
between L0 and T8, and sensitivity is also good at H band.
Sensitivity is lower at J and L bands, but useful work should still be
possible at these wavelengths for a smaller number of targets.

As part of this work I have also compiled all detections of
photometric variability, rotation periods, and \vsini\ for substellar
objects; these measurements are shown in Fig.~\ref{fig:prv} and listed
in Table~\ref{tab:database}.  Exceptionally bright and nearby brown
dwarfs (such as Luhman~16B), are feasible targets for Doppler Imaging
even on $<4$\,m telescopes. With 8\,m telescopes a few more objects
can be observed, but the number of accessible objects with measured
variability and/or rotation parameters are still small.

Doppler Imaging of substellar objects will hit its stride with the
next generation of giant ground-based telescopes, as shown in
Fig.~\ref{fig:sens}. Among objects already known, over 50 brown dwarfs
with spectral types $>$M9 should be amenable to this analysis. In
addition, Doppler Imaging will likely be possible for a small number
of young, self-luminous extrasolar giant planets.

The current sample of objects with measured variability, rotation,
and/or \vsini\ is dominated by early-type objects and becomes quite
sparse below $\sim$L5 (see Fig.~\ref{fig:prv}).  However, T dwarfs
occur more frequently in the Solar neighborhood than L dwarfs
\citep{kirkpatrick:2012}; this suggests that many additional targets
likely remain to be found.  There's plenty of room at the bottom of
the H-R diagram for new measurements of substellar rotation and
variability and, eventually, for Doppler Imaging.

\section*{Appendix: Data Sources}

\subsection*{Rotation Periods}
Individual references are listed in Table~\ref{tab:database}, and
include the following: \cite{artigau:2009}, \cite{bailerjones:2001},
\cite{berger:2005,berger:2009}, \cite{biller:2013},
\cite{buenzli:2012}, \cite{caballero:2004}, \cite{chew:2009},
\cite{clarke:2002a,clarke:2002b,clarke:2008}, \cite{enoch:2003},
\cite{gillon:2013}, \cite{gizis:2013}, \cite{hallinan:2008},
\cite{harding:2013a,harding:2013b}, \cite{heinze:2013},
\cite{irwin:2011}, \cite{joergens:2003}, \cite{khandrika:2013},
\cite{koen:2003,koen:2005a,koen:2006,koen:2013a}, \cite{lane:2007},
\cite{littlefair:2006}, \cite{martin:2006}, \cite{radigan:2012,radigan:2014},
\cite{rockenfeller:2006b}, \cite{scholz:2011}, \cite{wolszczan:2014},
and \cite{zapateroosorio:2003,zapateroosorio:2004}.


\subsection*{Projected rotation velocity}
Individual references for each object are listed in
Table~\ref{tab:database}. Measurements were taken from the following
references: \cite{bailerjones:2004}, \cite{berger:2008},
\cite{blake:2007}, \cite{bochanski:2011}, \cite{caballero:2004}, \cite{crossfield:2014a},
\cite{delburgo:2009}, \cite{deshpande:2012}, \cite{gizis:2013},
\cite{joergens:2001}, \cite{jones:2005}, \cite{konopacky:2012},
\cite{kurosawa:2006}, \cite{martin:1997},
\cite{mohanty:2003,mohanty:2005}, \cite{muzerolle:2003},
\cite{reid:2002}, \cite{reiners:2006a,reiners:2008,reiners:2010},
\cite{reiners:2007}, \cite{rice:2010}, \cite{snellen:2014}, \cite{tinney:1998},\cite{white:2003},
and \cite{zapateroosorio:2004,zapateroosorio:2006}.


\subsection*{Variability}
Individual references for measurements of significant photometric
and/or spectroscopic variability, including the relevant bandpasses,
are listed in Table~\ref{tab:database}. Data are taken from the
following sources: \cite{apai:2013}, \cite{artigau:2009},
\cite{bailerjones:2008}, \cite{bailerjones:2003},
\cite{bailerjones:1999,bailerjones:2001},
\cite{berger:2005,berger:2008,berger:2009}, \cite{bessell:1991},
\cite{biller:2013}, \cite{buenzli:2012,buenzli:2014},
\cite{burgasser:2014}, \cite{carpenter:2002}, \cite{chew:2009},
\cite{clarke:2002a,clarke:2002b,clarke:2008}, \cite{cody:2010},
\cite{crossfield:2014a}, \cite{enoch:2003}, \cite{gelino:2002},
\cite{gillon:2013}, \cite{girardin:2013}, \cite{gizis:2013},
\cite{goldman:2008}, \cite{harding:2013a,harding:2013b},
\cite{heinze:2013}, \cite{joergens:2003}, \cite{irwin:2011}, \cite{khandrika:2013},
\cite{koen:2003,koen:2004a,koen:2013a,koen:2005b,koen:2013b},
\cite{koen:2004b,koen:2005c}, \cite{lane:2007}, \cite{liebert:2003},
\cite{littlefair:2006,littlefair:2008}, \cite{maiti:2005,maiti:2007},
\cite{martin:2001}, \cite{radigan:2012,radigan:2014}, \cite{reid:2002},
\cite{reiners:2008}, \cite{rockenfeller:2006a,rockenfeller:2006b},
\cite{schmidt:2007,schmidt:2014},
\cite{scholz:2005,scholz:2006,scholz:2011}, \cite{stelzer:2006},
\cite{tinney:1999}, \cite{wilson:2014},
and \cite{zapateroosorio:2003,zapateroosorio:2004}. The interested reader
might also consult the work of \cite{ledesma:2009} to find a large
number of variable candidates ($>800$) in the Orion Nebula Cluster
(not included in this catalog).

\subsection*{Spectral type}
Spectral types are the NIR classifications from the DwarfArchives.org
website, from the literature, and from SIMBAD. Individual references
are listed in Table~\ref{tab:database}. For completeness, the
references used are: \cite{allen:2007},
\cite{artigau:2006,artigau:2010}, \cite{basri:2000},
\cite{bonnefoy:2013} \cite{bouy:2003},
\cite{burgasser:2007,burgasser:2006,burgasser:2008,burgasser:2013},
\cite{burningham:2010}, \cite{chiu:2006}, \cite{close:2003},
\cite{cruz:2003,cruz:2007,cruz:2009}, \cite{cushing:2011},
\cite{damjanov:2007}, \cite{deacon:2011}, \cite{fan:2000},
\cite{forveille:2005}, \cite{freed:2003},
\cite{gizis:2002,gizis:2000a,gizis:2000b,gizis:2001,gizis:2011},
\cite{goto:2002}, \cite{hawley:2002}, \cite{hewett:2006},
\cite{jenkins:2009}, \cite{kendall:2004,kendall:2007},
\cite{kirkpatrick:1995,kirkpatrick:1999,kirkpatrick:2000,kirkpatrick:2008,kirkpatrick:2010,kirkpatrick:2011},
\cite{knapp:2004}, \cite{kniazev:2013}, \cite{kraus:2007},
\cite{leggett:1994}, \cite{liebert:2003}, \cite{lodieu:2005},
\cite{luhman:2007}, \cite{martin:1999,martin:2000},
\cite{mohanty:2003,mohanty:2005}, \cite{phanbao:2008},
\cite{potter:2002}, \cite{reid:2001,reid:2006a,reid:2008},
\cite{reiners:2010}, \cite{rice:2010}, \cite{salim:2003},
\cite{schmidt:2007,schmidt:2014}, \cite{seifahrt:2010},
\cite{siegler:2003,siegler:2005}, \cite{stassun:2007},
\cite{thorstensen:2003}, \cite{tinney:1998,tinney:2005},
\cite{wilson:2003}, and \cite{zapateroosorio:2006}.

\subsection*{Binarity and Separations}

{\reff Reported binarity and measured separations are also included in the
electronic version of Table~\ref{tab:database}.  These are taken from
SIMBAD and the work of \cite{konopacky:2010}, \cite{luhman:2013},
\cite{martin:1999}, \cite{mccaughrean:2004}, \cite{probst:1983},
\cite{reid:2000}, \cite{reid:2001}, \cite{scholz:2010},
\cite{white:1999}, \cite{wilson:2014}, and \cite{zapateroosorio:2004}
}

\begin{acknowledgements}
  I thank V.\ J\"oergens for several excellent discussions that
  significantly improved the quality of the paper.  I thank I.\ Snellen
  for interesting and stimulating discussions during the preparation
  of this work, and for his subsequent suggestions while acting as
  referee. This research has made use of the TOPCAT software package,
  and of free and open-source software provided by the Python, SciPy,
  and Matplotlib communities. This research has benefited from the M,
  L, T, and Y dwarf compendium housed at DwarfArchives.org, and from
  SIMBAD.
\end{acknowledgements}

\clearpage

\begin{deluxetable}{l | r l l l l}
\tabletypesize{\small}
\tablecaption{Active, Planned, and Notional High-Resolution IR Spectrographs\label{tab:instruments}}
\tablewidth{0in}
\tablehead{

\colhead{Instrument} & \colhead{Resolution} & \colhead{Wavelength Coverage} & \colhead{Telescope} & Status & \colhead{References} 
}
\startdata

CSHELL &  40,000 & 1--5~$\mu$m, $\sim0.0025\lambda$ coverage & IRTF (3m) & Operating & \cite{greene:1993} \\
Phoenix & 70,000 & 1--5~$\mu$m, $\sim0.005\lambda$ coverage & KPNO (4m) & Operating & \cite{hinkle:1998}\\
ARIES &   \reff{50,000} & 1--2.5~$\mu$m simultaneous & MMT (6.5m) & Operating & \cite{mccarthy:1998}\\
CRIRES &   90,000 & 1--5~$\mu$m, $\sim0.02\lambda$ coverage\tablenotemark{a} & VLT (8m) & Operating\tablenotemark{a} & \cite{kaufl:2004}\\
IRCS   &   20,000 & 1--5~$\mu$m, $\sim0.2\mu m$ coverage & Subaru (8m) & Operating & \cite{tokunaga:1998} \\
NIRSPEC&   20,000 & 1--5~$\mu$m, $\sim0.1\lambda$ coverage & Keck (10m) & Operating & \cite{mclean:1998}\\ \hline
IGRINS &   40,000 & 1.5--2.5~$\mu$m simultaneous  & McDonald (2.7m) & Commissioning & \cite{yuk:2010}\\
GIANO  &   50,000 & 1--2.5~$\mu$m simultaneous    & TNG (3.6m) &  Commissioning & \cite{olivia:2012}\\
ISHELL &   72,000 & 1--5~$\mu$m, $\sim0.1\lambda$ coverage & IRTF (3m) & Under construction &  \cite{rayner:2012}\\
CARMENES & 82,000 & 0.6-1.7$\mu$m simultaneous & Calar Alto (3.5m) & Under construction & \cite{quirrenbach:2012} \\
SPIRou  & 75,000 & 1-2.4$\mu$m simultaneous & CFHT (3.6m) & Under construction & \cite{thibault:2012} \\
IRD    &   70,000 & 1--1.75~$\mu$m simultaneous   & Subaru (8m) &  Under construction & \cite{tamura:2012}\\
HPF    &   50,000 & 0.95--1.35~$\mu$m simultaneous& HET (9m) & Under construction & \cite{mahadevan:2012}    \\
{\reff HiJak}  &  {\reff 60,000} & {\reff 0.8--2.5 $\mu$m simultaneous} & {\reff DCT (4.3m)} & {\reff Notional} & {\reff Muirhead et al.\ (in prep.)}\\
iLocater& 100,000 & 0.95--1.1~$\mu$m simultaneous & LBT (8m)  & Planned & \cite{crepp:2014}\\
METIS  &  100,000 & 2.9--5.5~$\mu$m simultaneous  & E-ELT (39m) & Planned & \cite{brandl:2010,brandl:2012}\\
HIRES  &  100,000 & 0.4--2.5 $\mu$m simultaneous  & E-ELT (39m) &  Notional & \cite{maiolino:2013} \\
{\reff NIRES-B} & {\reff20,000} & {\reff 1--2.5 $\mu$m simultaneous} & {\reff TMT (30m)} & {\reff Notional} & {\reff\cite{TMTSRD:2013}} \\ 
{\reff NIRES-R} & {\reff100,000} & {\reff 3--5 $\mu$m simultaneous} & {\reff TMT (30m)} & {\reff Notional} & {\reff \cite{TMTSRD:2013}} \\ 
GMTNIRS&$\ge$60,000 & 1--5 $\mu$m simultaneous    & GMT (25m) &  Notional & \cite{jaffe:2006,lee:2010}\\
\enddata
\tablenotetext{a}{CRIRES is scheduled to be upgraded during 2014--2017 to provide $\sim0.2\lambda$ coverage. }
 \end{deluxetable}


\begin{deluxetable}{l l l r c c c c c }
\tabletypesize{\small}
\tablecaption{Catalog of Rotation and Variability of Substellar Objects\tablenotemark{a,b} \label{tab:database}}
\tablewidth{0in}
\tablehead{
\colhead{} & \colhead{RA} & \colhead{DEC}& \colhead{\vsini}& \colhead{$P$}& \colhead{$\Delta F$}& \colhead{}& \colhead{Spectral} & \colhead{Young?} \\
\colhead{Object} & \colhead{(hh mm ss.s)} & \colhead{($\pm$dd mm ss.s)}& \colhead{(\kms)}& \colhead{(hr)}& \colhead{(\%)}& \colhead{Bandpass}& \colhead{Type} & \colhead{(Y/N)}
}
\startdata
                 GJ 1001B &    00 04 34.8 &   -40 44 05.8 &           $34.4 \pm  1.8$ &   --- &         --- &         --- &    L4.5  &  N  \\
     2MASS 0019262+461407 &    00 19 26.2 &    +46 14 07 &           $68.0 \pm 10.0$ &   --- &         --- &         --- &    M8.0  &  N  \\
     2MASS 0019457+521317 &    00 19 45.7 &    +52 13 17 &           $ 9.0 \pm  2.0$ &   --- &         --- &         --- &    M9.0  &  N  \\
          DENIS 0021-4244 &  00 21 05.9 &  -42 44 43.33 &           $17.5 \pm  2.0$ &   --- &         --- &         --- &    M9.5  &  N  \\
     2MASS 0024246-015819 &  00 24 24.6 &  -01 58 20.14 &           $34.2 \pm  1.6$ &   --- &         1.6 &           I &    M9.5  &  N  \\
    2MASS 0024442-270825B &    00 24 44.2 &     -27 08 25 &           $ 9.0 \pm  2.0$ &   --- &         --- &         --- &    M8.5  &  N  \\
               LP 349-25A &   00 27 55.9 &   +22 19 32.8 &           $55.0 \pm  2.0$ &   --- &         --- &         --- &      M8  &  N  \\
               LP 349-25B &   00 27 55.9 &   +22 19 32.8 &           $83.0 \pm  3.0$ &   1.9 &         1.4 &           I &      M9  &  N  \\
... & ... & ... & ... & ... & ... & ... & ...  & ... \\
\enddata
\tablenotetext{a}{The full table is available online in
  machine-readable form. A portion is shown here as an example of its
  content.}
\tablenotetext{b}{References for all quantities are included in the machine-readable table, and for completeness are cited in the Appendix.}

 \end{deluxetable}

\clearpage 
\bibliographystyle{apj_hyperref}


\end{document}